\documentclass[jcs]{iosart2x}

\usepackage{lscape}
\usepackage{afterpage}

\usepackage{xcolor}

\usepackage{array}
\newcolumntype{V}{>{\centering\arraybackslash}p{0.1\linewidth}}

\newcommand{\rev}[1]{\textcolor{black}{#1}}

\pubyear{0000}
\volume{0}
\firstpage{1}
\lastpage{1}

\begin{document}

\makeatletter
\let\put@numberlines@box\relax
\makeatother

\begin{frontmatter}

\title{
Benchmarking Frameworks and Comparative Studies of Controller Area Network (CAN) Intrusion Detection Systems: A Review}
\runtitle{Review of CAN IDS Benchmarking Frameworks \& Comparative Studies}

\begin{aug}
    \author[A]{\inits{S.}\fnms{Shaila} \snm{Sharmin}\ead[label=e1]{shailasharmin@protonmail.com}%
    \thanks{Corresponding author. \printead{e1}.}}
    
    \author[A]{\inits{H.}\fnms{Hafizah} \snm{Mansor}\ead[label=e2]{hafizahmansor@iium.edu.my}}
    
    \author[A]{\inits{A.F.}\fnms{Andi Fitriah} \snm{Abdul Kadir}\ead[label=e3]{andifitriah@iium.edu.my}}

    \author[A]{\inits{N.A.}\fnms{Normaziah A.} \snm{Aziz}\ead[label=e4]{naa@iium.edu.my}}
    
    \address[A]{Kulliyyah of Information and Communication Technology, \orgname{International Islamic University Malaysia},
    Selangor, \cny{Malaysia}\printead[presep={\\}]{e1,e2,e3,e4}}
\end{aug}

\begin{abstract}
The development of intrusion detection systems (IDS) for the in-vehicle Controller Area Network (CAN) bus is one of the main efforts being taken to secure the in-vehicle network against various cyberattacks, which have the potential to cause vehicles to malfunction and result in dangerous accidents. These CAN IDS are evaluated in disparate experimental conditions that vary in terms of the workload used, the features used, the metrics reported, etc., which makes direct comparison difficult. Therefore, there have been several benchmarking frameworks and comparative studies designed to evaluate CAN IDS in similar experimental conditions to understand their relative performance and facilitate the selection of the best CAN IDS for implementation in automotive networks. This work provides a comprehensive survey of CAN IDS benchmarking frameworks and comparative studies in the current literature. A CAN IDS evaluation design space is also proposed in this work, which draws from the wider CAN IDS literature. This is not only expected to serve as a guide for designing CAN IDS evaluation experiments but is also used for categorizing current benchmarking efforts. The surveyed works have been discussed on the basis of the five aspects in the design space---namely IDS type, attack model, evaluation type, workload generation, and evaluation metrics---and recommendations for future work have been identified. 
\end{abstract}


\end{frontmatter}


\section{Introduction}\label{s1_introduction}

The adoption of drive-by-wire technology means that today's vehicles are equipped with as many as 150 Electronic Control Units (ECUs) \cite{charette_how_2021} which control the different subsystems of a vehicle and enable various functionalities related to performance, safety, and comfort. The operations of a vehicle rely on the communication of these ECUs among themselves, which occurs on the internal vehicular network connecting these ECUs together. One such protocol for internal vehicular networks is the Controller Area Network (CAN), which is used in nearly all modern vehicles due to its simple, inexpensive, and reliable implementation. However, the CAN bus also lacks security features, namely encryption and authentication, which makes it vulnerable to a range of attacks that can be conducted through any of the various communication interfaces a modern vehicle is equipped with, such as Bluetooth, Wi-Fi, and cellular.

Therefore, the need to secure the CAN bus has resulted in the development of CAN intrusion detection systems (IDS), particularly because a CAN IDS can be implemented without affecting the real-time performance of a CAN bus in the resource-constrained environment of an in-vehicle network \cite{agbaje_framework_2022, wu_survey_2020}. As such, there have been numerous intrusion detection methods proposed for the CAN bus over the years \cite{wu_survey_2020, aliwa_cyberattacks_2021, karopoulos_demystifying_2022}. 

These works provide results of evaluation experiments to demonstrate the efficacy of their proposed methods. But across the CAN IDS literature, these evaluation methods vary in terms of the workload used, the features used, the parameters chosen, and the metrics reported. Since the evaluation of these CAN IDS is performed with different experimental setups and in different contexts, it is not known how they fare in comparison to each other. Furthermore, the evaluation experiments reported in these works may not be replicable because the implementations are not readily available or documentation is not sufficiently comprehensive to aid reproduction \cite{agbaje_framework_2022, ji_comparative_2018}. This has resulted in efforts to evaluate CAN intrusion detection methods on an equal footing using benchmarking frameworks and in comparative studies. Evaluating CAN intrusion detection methods in similar settings facilitates the understanding of the relative performance of the CAN IDS under test and, ultimately, the selection of the best CAN IDS for implementation. 

Given the various ways in which intrusion detection methods can be evaluated \cite{milenkoski_evaluating_2015}, this work proposes an evaluation design space for the CAN IDS context that includes IDS types, attacks tested, evaluation type, workload source, and evaluation metrics. This design space enumerates the various evaluation methods found in the CAN IDS literature and is aimed at serving as a guide for planning future CAN IDS evaluation experiments. A survey of benchmarking and comparison studies of CAN IDS has also been conducted and the works discussed in terms of the proposed design space to understand current efforts at benchmarking and identify avenues for future work. The contributions of this paper can thus be summarized as follows: 

\begin{enumerate}
    \item Outlining a CAN IDS evaluation design space to aid the design of evaluation and benchmarking experiments.
    \item Providing a comprehensive survey of benchmarking frameworks and comparative studies of CAN IDS in the current literature, as well as categorizing and discussing the surveyed works according to the proposed design space.
    \item Discussing trends in current benchmarking and comparison efforts and providing recommendations for future work. 
\end{enumerate}

\rev{Contrary to prior survey works on CAN intrusion detection, this paper does not focus on surveying CAN IDS of different categories but rather on enumerating the various methods used for evaluating and benchmarking CAN IDS and organizing the information into an evaluation design space. We also survey benchmarking studies of CAN IDS, which have only been done for conventional IDS for computer networks. Table \ref{related_works} outlines how this work compares with related literature on CAN IDS and conventional IDS.}

The rest of this paper has been organized as follows. Section \ref{s2_background} provides background information on the CAN bus and relevant threats. Section \ref{s3_literature} provides an overview of related works in the literature, including the state of the art in CAN IDS and the benchmarking and evaluation of traditional computer networks. The proposed CAN IDS evaluation design space is described in Section \ref{s4_designspace}. Section \ref{s5_survey} presents the survey of benchmark frameworks and comparative studies of CAN IDS, while Section \ref{s6_discussion} discusses the surveyed works in terms of the five parts of the design space. Section \ref{s7_futurework} discusses opportunities for future work, while Section \ref{s8_conclusion} concludes this paper.  

\begin{table*}
\caption{\rev{Comparison of Present Study with Related Works}}
\resizebox{\columnwidth}{!} {%
\begin{tabular}{p{0.12\linewidth}VVVVVVV}
\hline
\textbf{Work}           				& \textbf{Focuses on CAN}	& \centering\textbf{Discusses IDS evaluation methods, datasets and/or evaluation metrics}	& \textbf{Proposes an IDS evaluation design space or framework} & \textbf{Surveys or updates previous surveys of IDS}	& \textbf{Surveys benchmarking studies of IDS}	& \textbf{Performs and reports new IDS benchmarking experiments}	& \textbf{Provides recommendations for future work} 	\\	\hline
Aliwa et al. \cite{aliwa_cyberattacks_2021}		& \checkmark     		&                                                                      		&                                                      		& \checkmark                                 		&                                     		&                                                       		& \checkmark                               		\\	\hline
Wu et al. \cite{wu_survey_2020}				& \checkmark     		& \checkmark                                                           		&                                                      		& \checkmark                                 		&                                    		&                                                       		& \checkmark                               \\	\hline
Karopoulos et al. \cite{karopoulos_demystifying_2022}	& \checkmark     		& \checkmark                                                          		&                                                      		& \checkmark                                 		&                                     		&                                                       		& \checkmark                               \\	\hline
Young et al. \cite{young_automotive_2019}		& \checkmark     		&                                                                     		&                                                      		& \checkmark                                 		&                                     		&                                                       		&                                          \\	\hline
Al-Jarrah et al. \cite{al-jarrah_intrusion_2019}	& \checkmark     		& \checkmark                                                           		&                                                      		& \checkmark                                 		&                                     		&                                                       		& \checkmark                               \\ 	\hline
Nappi \cite{nappi_survey_2022}				& \checkmark     		& \checkmark                                                           		&                                                      		& \checkmark                                 		&                                     		&                                                       		&                                          \\	\hline
Rajapaksha et al. \cite{rajapaksha_ai-based_2022}	& \checkmark     		& \checkmark                                                           		&                                                      		& \checkmark                                 		&                                     		&                                                       		& \checkmark                               \\	\hline
Panigrahi et al. \cite{panigrahi_performance_2021}	&                		& \checkmark                                                           		&                                                      		&                                            		& \checkmark                          		& \checkmark                                            		&                                          \\	\hline
Kilincer et al. \cite{kilincer_machine_2021}		&                		& \checkmark                                                           		&                                                      		&                                            		& \checkmark                          		& \checkmark                                            		& \checkmark                               \\	\hline
Milenkoski et al. \cite{milenkoski_evaluating_2015}	&                		& \checkmark                                                           		& \checkmark                                           		&                                            		&                                     		&                                                       		& \checkmark                               \\	\hline
\textbf{Our work}					& \textbf{\checkmark} 		& \textbf{\checkmark}                                                  		& \textbf{\checkmark}                                  		&                                            		& \textbf{\checkmark}                  		&                                                       		& \textbf{\checkmark}                      \\	\hline	
\end{tabular}%
}
\label{related_works}
\end{table*}

\section{Background}\label{s2_background}

\subsection{Controller Area Network}

CAN is a serial multi-master communication protocol and is one of the most commonly used for internal vehicular networks. It particularly finds use for subsystems such as powertrain and chassis, which are integral to the operation of a vehicle and include functionality such as transmission, braking, and steering \cite{aliwa_cyberattacks_2021}.

The CAN protocol spans the physical and data link layers of the OSI model. It is a message-based protocol whereby ECUs (i.e., nodes) communicate information related to the current state of the vehicle via message broadcasts that are received by all the other nodes on the network. A CAN message mainly consists of an arbitration identifier (AID) and a message payload of up to 8 bytes, along with other fields like the data length code (DLC) and cyclic redundancy check (CRC). The AID is used in the bit-wise arbitration process in the event of collision, whereby AIDs with lower values have higher priority. CAN also provides error checking and error confinement mechanisms. \cite{corrigan_introduction_2016}

\subsection{CAN attack model}\label{can_attack_model}

The CAN protocol was designed at a time when the internal vehicular network operated in isolation from external networks. Therefore, the protocol does not provide for security in its design and lacks encryption and authentication. Aspects of the CAN protocol, such as the arbitration mechanism, can also be exploited. These factors, along with the fact that modern vehicles are equipped with various interfaces for external communication, make the CAN bus vulnerable to a range of cyber attacks that can disrupt the operations of a vehicle and cause dangerous -- even fatal -- accidents.

Cho and Shin \cite{cho_fingerprinting_2016} provide an attack model for the CAN bus that is used in several works, such as \cite{verma_addressing_2020, islam_graph_2022}. This attack model assumes a compromised node on the CAN bus and uses the following terminology: a weakly compromised ECU is one that has been silenced or suspended by an attacker but cannot be used by the attacker to inject messages; on the other hand, a fully compromised ECU is one that the attacker has full control over and can use to inject messages into the CAN bus. This attack model classifies CAN bus attacks into the following categories: 

\begin{enumerate}
    \item \rev{\textbf{Fabrication attacks:}} These attacks involve the injection of fabricated messages into the CAN bus using a fully compromised ECU with the intention of overriding messages broadcast by particular ECUs or disrupting CAN bus communications. Fabrication attacks constitute the most common category of attacks in the CAN intrusion detection literature and include the following:
        \begin{enumerate}
            \item \rev{\textbf{Denial-of-service (DoS):}} \rev{A DoS attack can be carried out by injecting fabricated messages with the AID 0x000 or any AID lower in value than other legitimate AIDs, at a high frequency. Such messages would have the highest priority and always win the arbitration process, thereby blocking the broadcast of legitimate messages on the CAN bus.}
            

            \item \rev{\textbf{Fuzzing:}} Fuzzing involves the insertion of messages with random AIDs and payloads into the CAN bus at a high frequency. In some attacks the AIDs may be random, while in others the AID may be legitimate but the payload is random. 

            \item \rev{\textbf{Targeted ID:}} This type of attack involves the injection of messages of a particular AID with a manipulated payload. Such messages can be injected with a flooding delivery, i.e., at a high frequency, or with a flam delivery, where each forged message is injected immediately after a legitimate message of the same AID.

            \rev{Injection of messages that were previously seen and captured from the CAN bus, commonly termed as a \textbf{replay attack}, can also be considered a form of fabrication targeted ID attack. A replay attack can be conducted by injecting a previously captured sequence of messages, thereby targeting multiple AIDs, or by injecting a message time series of a single AID.}
        \end{enumerate}

    \item \rev{\textbf{Suspension attack:}} \rev{A suspension attack involves inhibiting message broadcasts from a weakly compromised ECU. This achieves an effect similar to a DoS and is observable as an absence of messages of a particular AID.}
    

    \item \rev{\textbf{Masquerade attack: }} Mounting a masquerade attack requires two main steps: the first is to suspend the message broadcasts from a weakly compromised ECU, and the second is use a fully compromised ECU to broadcast messages with the same AID as the former ECU but with a manipulated payload to effectively masquerade as the former ECU. 
\end{enumerate}

The attacks thus described range in their sophistication and in the methods that would be effective for detecting them. Attacks such as DoS and fuzzing do not require knowledge of the meaning and semantics of CAN broadcasts, which is often confidential, proprietary information and can vary among vehicle makes and models. They can also be detected by simpler methods, such as those that analyze the timing of AIDs. On the other hand, a masquerade attack is an advanced attack requiring greater skill to mount and would also require detection methods that analyze the payload of the messages \cite{verma_addressing_2020}.

\section{Related works}\label{s3_literature}

\subsection{CAN intrusion detection}\label{s3_1_canids}

Apart from the development of CAN encryption and authentication methods, the development of intrusion detection systems has been one of the main approaches being taken to secure the CAN bus \cite{aliwa_cyberattacks_2021}. This is because it is possible to implement a CAN IDS in the resource-constrained environment of an internal vehicular network without impacting CAN bus traffic or requiring changes to the CAN protocol \cite{agbaje_framework_2022, wu_survey_2020, karopoulos_demystifying_2022}.

A large number of CAN IDS have been proposed in the literature that vary in terms of the techniques used, features used, deployment location, etc. Numerous surveys of CAN intrusion detection methods provide categorisation for these CAN IDS, such as in \cite{young_automotive_2019, wu_survey_2020}. In their survey of cyberattacks and countermeasures for in-vehicle networks, Aliwa et al. \cite{aliwa_cyberattacks_2021} enumerate and review cryptographic methods and IDS for the CAN bus, categorising CAN IDS according to the approach used for intrusion detection. Karopoulos et al. \cite{karopoulos_demystifying_2022} compile a meta-taxonomy of CAN IDS, which provides a way to categorize CAN IDS based on deployment location, detection technique, network layer, and reaction type.  

Depending on the \rev{\textit{technique}} used for intrusion detection, we classify CAN IDS into the following categories (shown in Figure \ref{fig_benchmarking}), which are similar to those proposed in \cite{aliwa_cyberattacks_2021} and the classification by type proposed by Karopoulos et al. \cite{karopoulos_demystifying_2022} in their metataxonomy. For each category, we cite representative works and guide the reader to more detailed surveys of CAN IDS such as \cite{young_automotive_2019, wu_survey_2020, aliwa_cyberattacks_2021, al-jarrah_intrusion_2019}: 

\begin{enumerate}
    \item \rev{\textbf{Signature-based IDS:}} These IDS are knowledge-based or rule-based IDS that rely on a database of signatures for intrusion detection. This type of IDS is exemplified by that proposed by Studnia et al. \cite{studnia_language-based_2018}, which uses formal language theory to derive attack signatures from specifications of the network and ECU behaviour. Since the architecture of the in-vehicle network  remains mostly unchanged throughout a vehicle's lifespan, the expected behaviour of an in-vehicle network can be used to derive attack signatures. Such an approach is useful for detecting common, known attacks with low false positive and false negative rates. However, to ensure that the IDS is able to detect new attacks, the signature database needs to be regularly updated. Moreover, IDS relying on attack signature databases may not be capable of detecting novel, unknown attacks.

    \item \rev{\textbf{Anomaly-based IDS:}} These are behaviour-based methods that use patterns observed in normal CAN bus traffic and detect deviations from these normal patterns as attacks on the CAN bus. This is currently the most common approach for CAN intrusion detection, with 38 out of 41 CAN IDS works surveyed by Karopoulos et al. \cite{karopoulos_demystifying_2022} from the years 2020--2022 being of this type. The wide variety of anomaly-based CAN IDS can be further categorized into the following: 

    \begin{enumerate}
    
        \item \rev{\textbf{Statistical methods:}} These IDS use statistical methods to build a model of normal CAN network traffic and use them to detect anomalies. A number of CAN IDS take advantage of the fact that most CAN messages are broadcast at fixed regular frequencies. Timing-based IDS analyse time intervals between messages and raise an alert when the observed time interval deviates from the normal by a certain threshold, such as in \cite{moore_modeling_2017, song_intrusion_2016}. Young et al. \cite{young_automotive_2019} find that frequencies of messages, as opposed to timing, are better indicators of attacks like fabrication, with better accuracy and false positives. Message frequencies are also used in \cite{olufowobi_anomaly_2019} with an adaptive cumulative sum (CUSUM) algorithm and in \cite{bozdal_winds_2021} which uses wavelet analysis. However, while these IDS are suitable for detecting anomalies for periodic AIDs, they cannot be used for AIDs broadcast aperiodically.
    
        Apart from timing and frequency, other statistical CAN IDS use features such as AIDs and message payloads for intrusion detection. Marchetti and Stabili \cite{marchetti_anomaly_2017} identify that normal CAN bus traffic contains recurring sequences of AIDs, and the occurrence of any unusual transition between AIDs is indicative of an attack. This approach is similar to the graph-based approach taken in \cite{islam_graph_2022}, whereby a graph representing valid AID transitions is built from normal CAN traffic, and the chi-square test is applied to features derived from this graph for anomaly detection. The entropy of AIDs and message payloads has also been used as a means to detect anomalies in CAN traffic, in \cite{muter_entropy_2011, marchetti_evaluation_2016, baldini_application_2020}. In order to detect attacks involving manipulation of data fields, an IDS based on calculating Hamming distances between consecutive payloads of the same AID has also been proposed \cite{stabili_detecting_2017}, which is found to be effective against fuzzing attacks but not attacks involving the injection of a previously recorded message sequence (replay attack). 

        Statistical intrusion detection methods can be described to be effective in detecting only those attacks that impact the features being analyzed. As such, while timing- or frequency-based CAN IDS can be useful for detecting fabrication and suspension attacks, they may be unsuitable for masquerade attacks, which manifest as manipulated data fields. Such IDS are still lightweight approaches compared to machine learning methods and can be easily implemented in vehicle-grade ECUs for real-time intrusion detection. 
    
        \item \rev{\textbf{Machine Learning (ML) methods:}} These IDS use ML algorithms to build a model of CAN bus traffic and use it to detect attacks on the CAN bus. Traditional learning algorithms that have been applied to CAN intrusion detection include Support Vector Machine (SVM) and k-Nearest Neighbours (kNN) \cite{alshammari_classification_2018, refat_detecting_2022}, one-class SVM (OCSVM) \cite{avatefipour_intelligent_2019}, Isolation Forest \cite{sharmin_intrusion_2021}. Apart from these algorithms that learn "shallow" models, many works now apply deep learning techniques that can learn complex patterns in CAN traffic. These include Deep Neural Networks (DNN) \cite{zhang_intrusion_2019, fenzl_continuous_2020}, Long Short-Term Memory (LSTM) \cite{hanselmann_canet_2020, hossain_lstm-based_2020}, autoencoders \cite{longari_cannolo_2021, novikova_autoencoder_2022, kukkala_indra_2020}, Convolutional Neural Networks (CNN) \cite{thiruloga_tenet_2022, javed_canintelliids_2021}, and Generative Adversarial Networks (GAN) \cite{seo_gids_2018}. While these works utilize features derived from CAN message frames such as AID and data fields, Xun et al. \cite{xun_vehicleeids_2022} propose a CAN IDS using voltage signals of ECUs, whereby 14 time-domain features are used to train a deep support vector domain description (SVDD) model. 
    
        CAN intrusion detection has been treated as both a supervised and an unsupervised learning problem. While supervised learning methods require labelled datasets for training the detection models, unsupervised methods require only benign data with no attack traffic. Unsupervised methods can thus also detect novel, unknown attack types, as opposed to supervised learning methods, which can detect only the attack types they have been trained with. A detailed survey of ML-based CAN IDS can be found in \cite{rajapaksha_ai-based_2022}. 
    \end{enumerate}

    \item \rev{\textbf{Hybrid IDS}} A hybrid IDS is one that combines signature- and anomaly-based techniques for intrusion detection. These IDS are designed to combine the benefits of both types of CAN IDS: while a signature-based technique can efficiently detect common and known attacks with low false positives, anomaly-based methods enable the detection of novel attack types. The CAN IDS proposed by Zhang et al. \cite{zhang_hybrid_2022} is an example of this type of IDS that includes a whitelist of valid AIDs, a DLC validity check, a time interval-based detector, and a DNN detector implemented in that order. The DNN detector further uses AIDs, data field Hamming distances, entropy of data fields, and certain data field bytes as features, which were obtained from feature selection. Another hybrid IDS is CANova \cite{nichelini_canova_2023}, which uses a reverse engineering algorithm to extract signals from CAN messages and categorize messages, which is then used to select and apply appropriate detection modules depending on the category. This IDS includes static rules-based, timing-based, Hamming distance-based, Vector Auto-Regression-based, and Recurrent Neural Network (RNN) autoencoder-based modules. CAN IDS such as these can reduce detection times since known attacks are quickly detected by more lightweight methods, while also ensuring that novel attacks are caught using more computationally intensive methods. 
\end{enumerate}

CAN IDS can also be divided into two categories depending on the \rev{\textit{layer}} of the OSI model on which they operate: 

\begin{enumerate}
    \item \rev{\textbf{Data link layer IDS:}} \rev{These IDS apply any of the aforementioned techniques to features derived from CAN message frames, such as AIDs, DLCs, and payload data for intrusion detection. Among payload-based CAN IDS, there are those that analyse raw data bytes (such as \cite{stabili_detecting_2017}) and others that analyse signals derived from the payload bytes (such as \cite{longari_cannolo_2021}). This type of CAN IDS forms the majority of CAN IDS in the literature, as can be seen in the various surveys of CAN IDS \cite{wu_survey_2020, aliwa_cyberattacks_2021, young_automotive_2019, al-jarrah_intrusion_2019}.}
    

    \item \rev{\textbf{Physical layer IDS:}} These IDS use measurements of physical characteristics such as clock skew, voltage, and signal strength to perform intrusion detection. This type of IDS corresponds to OSI Layer 1. Cho and Shin \cite{cho_fingerprinting_2016} proposed a Clock-based IDS (CIDS), which estimates the clock skews of ECUs by analysing time intervals of periodic messages and uses them to build a profile of normal CAN traffic. On the other hand, VoltageIDS \cite{choi_voltageids_2018} and Viden \cite{cho_viden_2017} are CAN IDS that are based on using voltage measurements to fingerprint existing ECUs in a CAN bus and comparing observed CAN bus traffic with the fingerprints to detect intruding nodes. A survey of physical characteristics-based CAN IDS can be found in \cite{hafeez_state_2020}.
\end{enumerate}

As discussed in \cite{tomlinson_towards_2018}, the nature of the CAN in-vehicle network---characterized by limited computing resources, real-time response requirements, and a lack of sender and receiver identification---makes the design of CAN IDS different from those meant for computer networks. As such, a CAN IDS should ideally have low resource requirements, be able to detect and report attacks immediately, and be able to process the large number of CAN messages that are generated on the bus. In terms of attack detection accuracy, a CAN IDS should have a low false negative rate as well as a low false positive rate. Since CAN bus communications are used to control safety-critical subsystems in a vehicle, a low false negative rate is required to ensure that as many attacks as possible are detected. A low false positive rate is also required so that the CAN IDS does not raise an impractically large number of false alarms. ML-based IDS methods should also be designed to resist evasive, adversarial attacks. 

We find a larger number of works that utilize Layer 2 features like AIDs and data fields for intrusion detection, as opposed to physical characteristics-based CAN IDS, which are still a burgeoning area of study. Al-Jarrah et al. \cite{al-jarrah_intrusion_2019} also note the same in their review of in-vehicle intrusion detection systems where they examine surveyed works in terms of features and feature selection methods, datasets, performance metrics, benchmark models, and targeted attack types. They further find that most works focus on evaluating detection capability by reporting security-related metrics and do not report performance metrics like detection and training time. They also identify a lack of benchmark models and benchmark datasets for CAN IDS evaluation. Nappi \cite{nappi_survey_2022} updates their survey by reviewing more recent works, finding that more CAN IDS works now contextualize their findings through comparison with some benchmark models. However, there still remains a lack of standard benchmark models, and works that report detection latency still number in the minority.

\subsection{IDS benchmarking and comparison}

A brief overview of prior work in CAN IDS evaluation frameworks has been provided in \cite{agbaje_framework_2022}, whereby prior evaluation frameworks \cite{dupont_evaluation_2019, costa_canones_benchmarking_2021} have been reviewed to highlight key features and distinguish the proposed framework. Wu et al. \cite{wu_survey_2020} provide a description of datasets and tools used for evaluations in prior work and compare selected CAN IDS in terms of CAN IDS type, false positive rates, contributions, and drawbacks. However, apart from these, there have not been any comprehensive surveys of benchmarking and comparison efforts for CAN IDS, to the best of our knowledge. 

Similar reviews of comparative studies have been carried out for conventional computer network IDS, which mainly focus on the comparison of ML methods. Panigrahi et al. \cite{panigrahi_performance_2021} note that central to the research into the usage of ML for intrusion detection is the selection of the most appropriate classifier for building IDS. Therefore, they provide a review of comparative studies that examine supervised learning methods for network intrusion detection, summarising the classification models evaluated, datasets used, evaluation metrics reported, and findings of the studies. An analysis of a total of 54 classifiers from six categories of classification models has also been presented, with 13 metrics related to detection capability reported. Similarly, Almomani et al. \cite{Almomani_Machine_2021} also enumerate comparative studies of ML classifiers for network intrusion detection and present an evaluation of 10 supervised learning methods, reporting accuracy, precision, and F1-score. Finally, Kilincer et al. \cite{kilincer_machine_2021} conducted a survey of ML approaches to network intrusion detection by focusing on five datasets that are most commonly used for network IDS research. The authors note that network IDS studies are generally limited to a few datasets, examine only one or few classification methods, and examine only few attack types. To address these issues, classical ML models like SVM, kNN, and DT have been developed and evaluated using the five datasets as benchmark models. Results of experiments are compared against prior work utilizing the same datasets, thus contextualising past results. 

Conventional computer network intrusion detection literature is surveyed by Milenkoski et al.  \cite{milenkoski_evaluating_2015} to examine methods employed in the evaluation of IDS, and an evaluation design space is proposed in an effort to categorize these common methods. A design space is described by Baum et al. \cite{baum_mapping_2000} as a "multidimensional space of design choices," consisting of a set of relevant dimensions that can be used to classify and describe entities in a specific domain. The IDS evaluation design space proposed by Milenkoski et al. \cite{milenkoski_evaluating_2015} consists of three elements: workload, metrics, and measurement methodology. Workloads for evaluating IDS are classified as benign, malicious, and mixed, based on the presence of attacks in the workload. The authors also distinguish between workloads in executable forms for live testing of CAN IDS and trace forms for later replay. Metrics are classified as security-related and performance-related metrics. The measurement methodology part of the design space identifies IDS properties that are of interest and the workload and metrics that are employed to evaluate these properties. This evaluation design space not only serves as a basis for categorising the literature, but is also aimed at facilitating the planning of IDS evaluation exercises.

The present work thus attempts to propose a design space for the context of CAN IDS evaluation, not only to categorize current comparative evaluation methods but also to provide a guideline for designing IDS evaluations for CAN, which differs from conventional computer networks in features and complexity. 

\section{CAN IDS evaluation design space}\label{s4_designspace}

\rev{Similar to the guidelines for evaluating conventional CAN IDS proposed in \cite{milenkoski_evaluating_2015}, we understand that planning a CAN IDS evaluation study should begin with identification of the goals of the study and the associated constraints. The goal of an evaluation study is usually the examination of the detection capability and/or the performance aspects of one or several CAN IDS. On the other hand, the main constraints in the evaluation of CAN IDS are the availability of resources such as vehicles, testbeds, and other tools, as well as the access to requisite data and information such as proprietary CAN database files that contain rules to decode CAN messages. Consideration of these goals and constraints should then inform the design of the evaluation study.}

\begin{figure*}[t]
    \includegraphics[width=\textwidth]{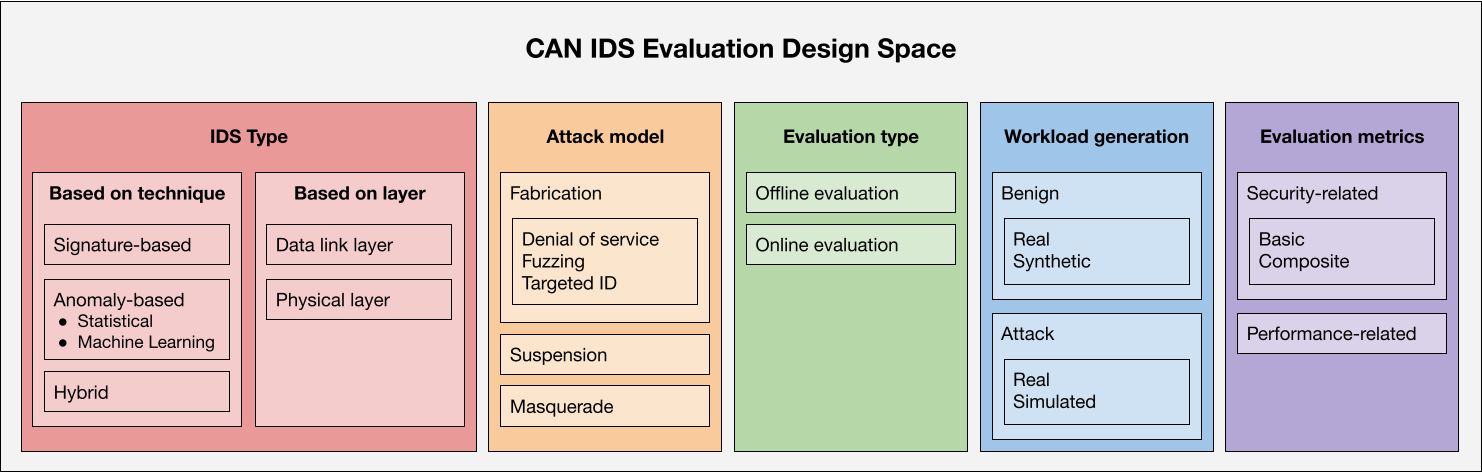}
    \caption{CAN IDS Evaluation Design Space}
    \label{fig_benchmarking}
\end{figure*}

\rev{Our proposed CAN IDS evaluation design space is not only aimed at enumerating the various evaluation methods available but is also meant to provide a way to describe any CAN IDS evaluation study completely. To do so, we have divided the design space into five essential components, summarized in Figure \ref{fig_benchmarking}: the IDS types being evaluated, the attacks considered, the evaluation type, the workload being used, and the evaluation metrics being reported. The choices that should be made for each component thus depend on the goals and constraints of the study, as well as on the choices made for related components. Choosing to report the accuracy, precision, and recall metrics to describe detection capability is an example of a goal influencing the evaluation metric choice; while choosing online tests on a lower-cost testbed instead of a real vehicle represents a choice of evaluation type resulting from a constraint. The decision to include only fabrication and suspension attacks for testing timing-based CAN IDS is a further example of the choice for one component (CAN IDS type) influencing that for another component (attack types).}

\rev{The remainder of this section describes in further detail each component of the CAN IDS evaluation design space and how consideration of goals, constraints, and related components influences the choices for each component.}


\subsection{IDS type}\label{designspace_idstype}

As described in Section \ref{s3_1_canids}, a CAN IDS can be either signature-based, anomaly-based, or hybrid, depending on the detection technique used. Anomaly-based methods are further divided into statistical and ML-based methods. Depending on the features used and the OSI layer they operate on, CAN IDS can either be a data link layer or a physical layer IDS. 

The classification of a particular CAN IDS informs the type of data or workload required for its evaluation---while a physical layer CAN IDS would require Layer 1 data, a data link layer CAN IDS would require logs containing CAN message frames. As a further example, a CAN IDS based on analysis of AID sequences would require only the AIDs of CAN broadcasts, while a CAN IDS that incorporates timing-based features would require high precision message timestamps. 

The detection technique employed by a given CAN IDS also informs the types of attacks that can be detected by the IDS and are relevant for inclusion in its assessment. This is because different attacks manifest as changes in different features of CAN bus traffic, and CAN IDS differ in the features used. Timing and frequency-based CAN IDS \cite{young_automotive_2019, moore_modeling_2017, bozdal_winds_2021, olufowobi_anomaly_2019, song_intrusion_2016} can be good at detecting fabrication and suspension attacks, which alter the timing and frequencies of AIDs, but ineffective for masquerade attacks, which are observed as manipulated payloads. Masquerade attacks can instead be detected by CAN IDS that analyse the data fields of CAN messages, such as the Hamming distance CAN IDS \cite{stabili_detecting_2017} or ML-based CAN IDS \cite{longari_cannolo_2021}. Physical characteristics-based CAN IDS, which fingerprint CAN IDS using physical features that are difficult to spoof, are capable of identifying malicious nodes and can thus not only detect fabrication and suspension attacks but also masquerade attacks \cite{cho_fingerprinting_2016, schell_valid_2020}.

\subsection{Evaluation type}

This part of the CAN IDS evaluation design space identifies two types of evaluations: offline and online evaluations. In an offline evaluation, a CAN IDS is used to analyse a CAN bus log or dataset that has already been collected from a real or simulated CAN bus. This is opposed to an online assessment, where a CAN IDS performs real-time analysis of CAN bus data from a real vehicular CAN bus, testbed, simulation, or data log replay. The physical characterstic-based CAN IDS proposed in \rev{\cite{cho_viden_2017, cho_fingerprinting_2016}} are evaluated in CAN bus prototypes with nodes consisting of Arduino boards and CAN shields, and with a real vehicle whereby the IDS is implemented on a node connected to the CAN bus via the OBD-II port. Ujiie et al. \cite{ujiie_method_2016} implement their rule-based CAN IDS on an ATMega162 microcontroller and test it against both a simulated CAN bus in Vector CANoe as well as with a real vehicle. Unlike these works, Desta et al. \cite{desta_id_2020} evaluate their CAN IDS, which uses an LSTM model for AID sequence prediction, by replaying a CAN bus data log and using the SocketCAN API to perform detection. 

Using real vehicles is advantageous in that they most closely resemble the real-world environment in which an in-vehicle CAN IDS would operate. But it is relatively difficult to use real vehicles for CAN IDS assessments.  Not only is it expensive to acquire and use a real vehicle for security testing, but mounting attacks like targeted ID, suspension, and masquerade on a real vehicular CAN bus can be difficult, time-consuming, and also pose a risk to passengers and bystanders \cite{verma_addressing_2020, rathore_invehicle_2022}. To address these problems, a testbed for the purpose of online CAN IDS evaluation has been developed in \cite{jadidbonab_real-time_2021}, which uses CARLA car simulator in combination with the Vector CANoe CAN bus simulator to generate realistic driving scenarios. The assessment of a clustering-based ML CAN IDS demonstrated inferior detection performance in the online assessment using the testbed compared to an offline experiment with a dataset collected from the same testbed, which highlights the importance of online assessment regardless of its drawbacks. 

On the other hand, offline assessments with collected CAN bus logs (as well as online tests by replaying logs) can be performed repeatedly with relative ease to obtain statistically significant evaluation results.  A large number of works, such as \cite{bozdal_winds_2021, longari_cannolo_2021, song_intrusion_2016, stabili_detecting_2017, marchetti_evaluation_2016, nichelini_canova_2023, zhang_hybrid_2022}, use collected CAN bus logs to assess their proposed CAN IDS. These datasets are commonly collected either via the OBD-II port available in all vehicles or by tapping into the in-vehicle CAN bus. Offline evaluation is a good starting point to understand the detection capability in terms of accuracy, false positive rates, etc. of a CAN IDS before using online evaluations to understand performance aspects of the CAN IDS such as detection times. Using publicly available datasets further enhances the reproducibility of CAN IDS works and allows direct comparison with results from other CAN IDS assessments performed with the same datasets.

\subsection{Workload}

Workload can be described as the work that must be performed by a system and can be viewed as input to the system \cite{gadelrab_new_2012}. In the context of a CAN IDS, its workload comes from CAN bus traffic or measurements of physical characteristics. \rev{While CAN IDS datasets serve as the most common source of data to evaluate CAN IDS, in this paper we borrow the term 'workloads' from the wider conventional IDS literature to describe any input to a CAN IDS under test regardless of the source it originates from---which could be not just datasets but real-time CAN bus traffic or measurements of physical quantities from a vehicle, testbed, or simulation in an online test.}

\begin{figure*}[t]
    \includegraphics[width=\textwidth]{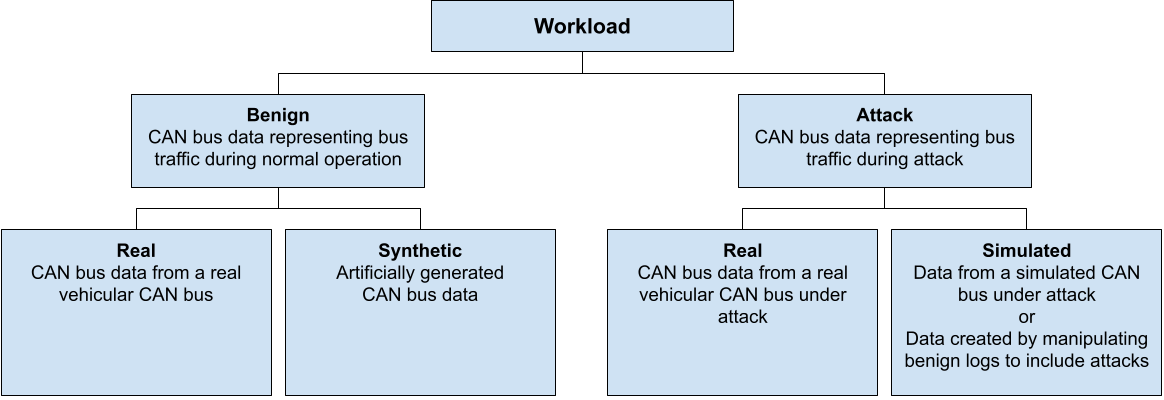}
    \caption{CAN IDS Workload Types}
    \label{fig_workload}
\end{figure*}

The workload used to evaluate an IDS may be generated in various ways \cite{milenkoski_evaluating_2015}. For CAN IDS evaluation, we may distinguish between benign, attack-free \rev{workloads} and malicious \rev{workloads containing attacks}. Benign \rev{CAN IDS workloads} may be from a real vehicle (real) or artificially generated (synthetic). Attack \rev{workloads} can be obtained by conducting attacks on a real CAN bus (real attacks) or via simulation, which can be by manipulating a collected benign CAN trace to include attacks or by artificially generating \rev{malicious CAN workloads} (simulated attacks). \rev{These various types of workloads are outlined in Figure \ref{fig_workload}.} 


A number of publicly available CAN datasets have been published in recent years, which has made them a popular choice for the design and evaluation of CAN IDS \cite{nappi_survey_2022}. The Hacking and Countermeasures Research Lab (HCRL) has published three CAN intrusion datasets \cite{lee_otids_2017, han_anomaly_2018, seo_gids_2018}, which contain benign CAN bus logs as well as logs of fabrication attacks such as DoS, fuzzing, and targeted ID conducted on a real in-vehicle CAN bus. These datasets have been used in \cite{paul_artificial_2021, islam_graph_2022, javed_canintelliids_2021, sharmin_intrusion_2021, khandelwal_lightweight_2022, barletta_intrusion_2020} among others. Verma et al. \cite{verma_addressing_2020}, who also provide a comprehensive survey of CAN intrusion datasets, present the Real ORNL Dynamometer (ROAD) dataset, consisting of real benign samples and samples of fuzzing, targeted ID (flam delivery), and masquerade attacks. While the fabrication attacks were conducted on a real vehicle CAN bus, the masquerade attack samples were created by manipulating the targeted ID logs. This dataset has been used in \cite{moriano_detecting_2022}. 

The dataset provided in \cite{dupont_automotive_2019} consists of logs collected from both a CAN bus prototype as well as real vehicles. Unlike the datasets mentioned thus far, the attack samples are simulated in that the benign data from vehicles has been augmented to create attack datasets (with the exception of the attack samples collected from the prototype). This dataset has been used in works such as \cite{sharmin_using_2022, bozdal_winds_2021}. In a similar vein, the CrySyS lab has published real benign CAN bus logs and provided a log infector tool that can be used to manipulate benign logs to create masquerade attack samples. This tool has been used to create the attack samples used to evaluate the statistical CAN IDS in \cite{gazdag_correlation-based_2022}. A completely synthetic dataset, SynCAN, with simulated targeted ID, suspension, and masquerade attacks is published by the authors of \cite{hanselmann_canet_2020}, where they use it to evaluate their LSTM autoencoder-based CAN IDS. It has also been used in \cite{novikova_autoencoder_2022, kukkala_indra_2020}, which are also autoencoder-based CAN IDS. Apart from ROAD, SynCAN is the only dataset providing translated signal values instead of raw CAN data fields.

Physical fingerprinting-based CAN IDS, unlike data link layer CAN IDS, cannot be evaluated using the aforementioned datasets. Towards this end, Foruhandeh et al. \cite{foruhandeh_simple_2019} have published a dataset consisting of voltage measurements from a real vehicle along with their Single-frame based Physical-Layer (SIMPLE) identification solution that can detect intrusions and identify the sending ECU for each message. This dataset is also used in \cite{lalouani_mitigating_2022} for the detection of a hill-climbing style masquerade attack whereby an attacker attempts to evade detection and alter ECU fingerprints to do so. Popa et al. \cite{popa_ecuprintphysical_2022} have also made available clock skew and voltage data from a total of 54 ECUs across 10 vehicles, which can be used for the development of physical layer CAN IDS.

\subsection{Attack model}

The attack model considered in this design space is the same as the one described in Section \ref{can_attack_model}. 

As discussed in section \ref{designspace_idstype}, the attack types that can be detected by a CAN IDS depend on the features and techniques used by the CAN IDS. However, this is not the only consideration a researcher has to make when selecting attack types for CAN IDS assessment. Fabrication attacks are relatively less complex compared to suspension and masquerade attacks and form the most common class of attacks found in the CAN IDS literature \cite{verma_addressing_2020}, which is reflected in the fact that real attack datasets are available for only these attack types \cite{verma_addressing_2020, han_anomaly_2018, seo_gids_2018, lee_otids_2017}. On the contrary, while CAN intrusion datasets with suspension and masquerade attacks are available, these are simulated attack samples created either from a purely synthetic benign dataset \cite{hanselmann_canet_2020} or a real benign dataset \cite{dupont_automotive_2019}. The advantage of using real attacks over simulated attacks is that in the former case, the attacks are known to have caused an effect on the operations of the vehicle, i.e., the effects are physically verified. For example, while creating the ROAD dataset, the authors of \cite{verma_addressing_2020} noted abnormal behaviour such as accelerator pedals becoming ineffective, false displays on speedometer, and incorrect reverse light status. With simulated attacks, not only is it impossible to verify their physical effects, but the attack simulation method (such as manipulating benign CAN bus logs) may result in an unrealistic attack sample. 

Another important consideration in selecting attack datasets for evaluation is the "difficulty" of the detection problem captured in the dataset \cite{vahidi_systematic_2022}. Verma et al. \cite{verma_addressing_2020} find that datasets such as \cite{lee_otids_2017} and \cite{seo_gids_2018} that are commonly used consist of unstealthy attacks with high frequency injection of malicious messages that can be detected by trivial, timing-based detectors and are less suitable for assessing more sophisticated CAN IDS that should be able to detect low rate fabrication attacks \cite{swessi_comparative_review_datasets_2021, verma_addressing_2020}. 

Since the usage of real test vehicles is outside the reach of many researchers, there is a need for comprehensive CAN intrusion datasets consisting of real attack traces of all known attack types, ranging from simple, easily detected attacks to complex, stealthy attacks, for robust CAN IDS evaluation and benchmarking. This has become necessary for physical layer CAN IDS as well, most of which continue to be evaluated with real vehicles and testbeds.

\subsection{Evaluation metrics}\label{designspace_evaluationmetrics}

As per \cite{milenkoski_evaluating_2015}, the metrics selected and reported while evaluating an IDS should depend on the properties of the IDS being assessed. Metrics can be security-related metrics, which quantify attack detection capability, or performance-based metrics, which quantify non-functional aspects of an IDS such as resource consumption. The authors of \cite{milenkoski_evaluating_2015} also distinguish between basic security metrics, which quantify individual attack detection properties, and composite security metrics, which combine basic metrics. 

Security-related metrics allow assessment of attack detection properties such as attack detection accuracy and resistance to evasive attacks. These include  classification accuracy, precision, recall, F1-score, and false positive rate, which have been reported in \cite{alshammari_classification_2018, song_intrusion_2016, young_automotive_2019, desta_id_2020, kukkala_indra_2020, zhang_hybrid_2022} among others. Basic security metrics such as true positive rate, false positive rate, false negative rate, true negative rate, precision, and recall must be reported and analyzed together to understand the performance of a given CAN IDS \cite{milenkoski_evaluating_2015}. Some works choose to report receiver operator characteristics (ROC) and area under curve (AUC) \cite{olufowobi_anomaly_2019, longari_cannolo_2021}, which are considered composite security metrics, to indicate the detection performance of an IDS at multiple operating points. 

An important problem to consider when using CAN datasets for assessment is the class imbalance in such datasets, whereby malicious messages that are a part of attack traffic are usually present as a very small percentage of total captured CAN bus traffic. This is further reason to not rely on a single metric like accuracy, which would yield high values for a detector that only predicts the normal class for a highly imbalanced dataset, and instead use a suite of security metrics to understand the ability of the IDS to distinguish between normal and attack traffic. While some may choose to counter the imbalanced dataset problem by reporting balanced accuracy, Chicco et al. \cite{chicco_matthews_2021} recommend use of the Matthews Correlation Coefficient (MCC), which has been described as a single metric that summarizes the performance of a binary detector. MCC is especially suited for CAN intrusion detection since it is equally important for a CAN IDS to correctly classify both normal and attack traffic, i.e., to keep both false positive and false negative rates low. The MCC is reported alongside other security metrics in \cite{nichelini_canova_2023}.

Considering the safety-critical nature and real-time requirements of the in-vehicle network, it may be argued that attack detection latency is an important metric that should be considered in assessing the attack detection capabilities of a CAN IDS \cite{corbett_testing_2017, nappi_survey_2022, al-jarrah_intrusion_2019}. Detection latency, or Time To Detection (TTD), has been defined as the time taken to classify a CAN message from the time it was received \cite{nappi_survey_2022} and has been reported in \cite{olufowobi_saiducant_2020}. Nichelini et al. \cite{nichelini_canova_2023} report Testing Time per Packet (TTP) as the ratio of the total detection time and the number of messages in their test dataset, which gives the average time taken by the CAN IDS to evaluate a single CAN message. Unlike these works, Sunny et al. \cite{sunny_hybrid_2020} report best case and worst case computation time to examine if their proposed CAN IDS can be used for real-time evaluation of CAN messages, which can be published every 2 ms. 

Apart from detection latency, non-functional properties of a CAN IDS like resource consumption, performance overhead, and workload processing capacity are also of interest, particularly since they are expected to be deployed in the resource-constrained environment of the in-vehicle network. A CAN IDS that is highly accurate in detecting attacks may still become impractical for implementation in the in-vehicle network if it is not able to process rapidly generated CAN bus traffic in time or requires significant computing resources to maintain quick response times. An analysis of computational complexity and memory requirement have been provided for the Hamming distance-based CAN IDS in \cite{stabili_detecting_2017}. Unlike works with offline assessment using datasets, memory footprint in kilobytes and inference time (similar to TTD) of the autoencoder-based detector in \cite{kukkala_indra_2020} have been measured by implementing the IDS on an automotive-grade microcontroller.

\section{Survey of benchmark frameworks and comparative studies of CAN IDS}\label{s5_survey}

This section provides an overview of the evaluation frameworks and comparative studies of CAN IDS that have been detailed in the literature. The papers included in this study were published between 2017 and 2022 (September) and were selected by conducting a search on Google Scholar, IEEEXplore, and the ACM Digital Library using the keywords "controller area network intrusion detection system benchmark", "controller area network intrusion detection system evaluation", "controller area network intrusion detection system testbed", "controller area network intrusion detection system comparative". Papers were included in and excluded from this study by considering abstracts. 

The surveyed works differ in their scope in terms of the types of IDS evaluated, the attack types tested, and the metrics reported. Since the works mostly restrict themselves to particular types of CAN IDS, they have been categorized as those that evaluate statistical CAN IDS (listed in Table \ref{table-stat-ids}) and those that evaluate ML-based CAN IDS (listed in Table \ref{table-ml-ids}). The attack types that have been used for evaluation by the surveyed works are summarized in Table \ref{table-attacks}, while the reported metrics are provided in Table \ref{table-metrics}.

\begin{table*}[]
\caption{Statistical CAN IDS Evaluated in Surveyed CAN IDS Benchmark Frameworks and Comparative Studies}
\resizebox{\columnwidth}{!}{%
\begin{tabular}{@{}llccc@{}}
\hline
\textbf{Work}							&\textbf{IDS Evaluated}                                                        				& \multicolumn{3}{c}{\textbf{Key Features Used}}	\\ \cline{3-5}
									&																	& \textbf{Timestamp}& \textbf{AID}	& \textbf{Payload}\\ \hline
Ji et al. (2018) \cite{ji_comparative_2018}		& Entropy-based \cite{muter_entropy_2011}                                                            	& 			& \checkmark	& 			\\
									& Clock skew-based \cite{cho_fingerprinting_2016}									& \checkmark	&			&			\\
									& ID sequence algorithm \cite{marchetti_anomaly_2017}                                                	& 			& \checkmark	& 			\\
									& Frequency-based \cite{taylor_frequency_2015}                                                       	& \checkmark	&			&			\\ \hline
Dupont et al. (2019) \cite{dupont_evaluation_2019}	& Diagnostic messages detection \cite{miller_survey_2014}                                            	& 			& \checkmark	& 			\\
									& Pattern matching$^{\mathrm{a}}$ \cite{ujiie_method_2016, abbott_mccune_intrusion_2016}                           	& 			& \checkmark	& \checkmark	\\
									& Time interval-based \cite{gmiden_intrusion_2016}                                                   	& \checkmark	&			&			\\
									& Frequency-based \cite{taylor_frequency_2015}                                                       	& \checkmark	&			&			\\
									& Time interval-based \cite{moore_modeling_2017}                                                     	& \checkmark	&			&			\\
									& Time interval-based \cite{song_intrusion_2016}                               				& \checkmark	&			&			\\
									& Entropy-based (for CAN message windows) \cite{marchetti_evaluation_2016, muter_entropy_2011}		& 			& \checkmark	& 			\\
									& Entropy-based (for flows of individual AIDs) \cite{marchetti_evaluation_2016, muter_entropy_2011}	& 			& \checkmark	& 			\\ \hline
Stabili et al. (2021) \cite{stabili_benchmark_2021}	& ID sequences algorithm \cite{marchetti_anomaly_2017}                                               	& 			& \checkmark	& 			\\
									& Entropy-based algorithm \cite{marchetti_evaluation_2016}                                           	& 			& \checkmark	& 			\\
									& Hamming distance \cite{stabili_detecting_2017}                                                     	&			&			& \checkmark	\\
									& Missing message algorithm \cite{stabili_detection_2019}								& \checkmark	&			&			\\ \hline
Agbaje et al. (2022) \cite{agbaje_framework_2022}	& Time interval-based \cite{moore_modeling_2017}                                                     	& \checkmark	&			&			\\
									& Frequency-based \cite{young_automotive_2019}                                                       	& \checkmark	&			&			\\
									& CUSUM \cite{olufowobi_anomaly_2019}                                                                	& 			& \checkmark	& 			\\
									& Entropy-based \cite{muter_entropy_2011}                                                            	& 			& \checkmark	& 			\\
									& Graph-based \cite{islam_graph_2022}                                                                	& 			& \checkmark	& 			\\
									& ID sequences algorithm \cite{marchetti_anomaly_2017}                                               	& 			& \checkmark	& 			\\
									& Hamming distance \cite{stabili_detecting_2017}									&			&			& \checkmark	\\
									& Neural network$^{\mathrm{b}}$ \cite{paul_artificial_2021}								& 			& \checkmark	& \checkmark	\\ \hline
Blevins et al. (2021) \cite{blevins_time-based_2021}	& Mean Inter-message time													& \checkmark	&			&			\\
									& Binning                                                                                            	& \checkmark	&			&			\\
									& Fitting a Gaussian curve                                                                           	& \checkmark	&			&			\\
									& Kernel Density Estimation                                                                          	& \checkmark	&			&			\\ \hline
Stachowski et al. (2019) \cite{stachowski_assessment_2019}$^{\mathrm{c}}$ & Anomaly-based IDS											& 			&			&			\\ \hline
\multicolumn{5}{l}{$^{\mathrm{a}}$Signature-based CAN IDS}			\\
\multicolumn{5}{l}{$^{\mathrm{b}}$ML-based CAN IDS}			\\
\multicolumn{5}{l}{$^{\mathrm{c}}$Details of CAN IDS evaluated are not disclosed}			
\end{tabular}%
}
\label{table-stat-ids}
\end{table*}

\begin{table*}[]
\caption{ML-Based CAN IDS Evaluated in Surveyed CAN IDS Benchmark Frameworks and Comparative Studies}
\resizebox{\columnwidth}{!}{%
\begin{tabular}{@{}llccc@{}}
\hline	
\textbf{Work}											&\textbf{IDS Evaluated}             & \multicolumn{3}{c}{\textbf{Key Features Used}}	\\ \cline{3-5}
													&						& \textbf{Timestamp}& \textbf{AID}	& \textbf{Payload}\\ \hline
Taylor et al. (2018) \cite{taylor_probing_2018}             			& Long Short-Term Memory network	&			&			& \checkmark	\\
                                       							& Gated Recurrent Unit network	&			&			& \checkmark	\\
                                       							& Markov chains				&			&			& \checkmark	\\ \hline
Berger et al. (2019) \cite{berger_comparative_2019}         			& One-Class Support Vector Machine	& \checkmark	& \checkmark	& \checkmark	\\
                                      							& Support Vector Machine		& \checkmark	& \checkmark	& \checkmark	\\
                                       							& Neural network				& \checkmark	& \checkmark	& \checkmark	\\
                                       							& Long Short-Term Memory network	& \checkmark	& \checkmark	& \checkmark	\\ \hline
Moulahi et al. (2021) \cite{moulahi_comparative_2021}$^{\mathrm{a}}$		& Support Vector Machine		& \checkmark	& \checkmark	& \checkmark	\\
                                       							& Decision Trees				& \checkmark	& \checkmark	& \checkmark	\\
                                       							& Random Forest				& \checkmark	& \checkmark	& \checkmark	\\
                                       							& Multi-layer perceptron		& \checkmark	& \checkmark	& \checkmark	\\ \hline
Costa Cañones (2021) \cite{costa_canones_benchmarking_2021} 			& Isolation Forest			&			&			& \checkmark	\\
                                       							& One-Class Support Vector Machine 	&			&			& \checkmark	\\
                                       							& Autoencoder with neural network	&			&			& \checkmark	\\
                                       							& Autoencoder with LSTM			&			&			& \checkmark	\\
                                       							& Autoencoder with GRU			&			&			& \checkmark	\\
                                       							& CANnolo (26)				&			&			& \checkmark	\\ \hline
Swessi and Idoudi (2021) \cite{swessi_comparative_2022}         			& Decision Trees				& \checkmark	& \checkmark	& \checkmark	\\ 
                                       							& Random Forest				& \checkmark	& \checkmark	& \checkmark	\\
                                       							& Bagging Tree				& \checkmark	& \checkmark	& \checkmark	\\
                                       							& Extra Trees				& \checkmark	& \checkmark	& \checkmark	\\
                                       							& Gradient Boosting			& \checkmark	& \checkmark	& \checkmark	\\
                                       							& Adaptive Boosting			& \checkmark	& \checkmark	& \checkmark	\\
                                       							& Voting					& \checkmark	& \checkmark	& \checkmark	\\
                                       							& Stacking					& \checkmark	& \checkmark	& \checkmark	\\
                                       							& eXtreme Gradient Boosting   	& \checkmark	& \checkmark	& \checkmark	\\
                                       							& Light Gradient Boosting     	& \checkmark	& \checkmark	& \checkmark	\\
                                       							& Category Gradient Boosting		& \checkmark	& \checkmark	& \checkmark	\\ \hline
Anyanwu et al. (2021) \cite{anyanwu_countering_2021}					& Tree                        	&			& \checkmark	& \checkmark	\\
                                       							& Support Vector Machine      	&			& \checkmark	& \checkmark	\\
                                       							& Ensemble Learning			&			& \checkmark	& \checkmark	\\
                                       							& Discriminant models         	&			& \checkmark	& \checkmark	\\
                                       							& Nearest Neighbour			&			& \checkmark	& \checkmark	\\
													& Logistic Regression			&			& \checkmark	& \checkmark	\\ \hline
Okokpujie et al. (2021) \cite{okokpujie_anomaly-based_2021}$^{\mathrm{a}}$	& Feedforward neural network  	&			& \checkmark	& \checkmark	\\
                                      							& Support Vector Machine		&			& \checkmark	& \checkmark	\\ \hline
\multicolumn{5}{l}{$^{\mathrm{a}}$These studies model intrusion detection as a multi-class classification problem where the attack type is predicted} 			\\ 
\end{tabular}%
}
\label{table-ml-ids}
\end{table*}

\afterpage{
\begin{landscape}

\begin{table}[]
\caption{Workloads Used and Attack Types Covered in Surveyed CAN IDS Benchmark Frameworks and Comparative Studies}
\resizebox{\linewidth}{!}{%
\begin{tabular}{@{}lp{0.25\linewidth}ccccccp{0.12\linewidth}@{}}
\hline
\textbf{Work}								& \textbf{Workload}                                                    						& \textbf{Publicly}	& \multicolumn{3}{c}{\textbf{Fabrication}}				& \textbf{Suspension} & \textbf{Masquerade} & \textbf{Other}                       	\\ \cline{4-6}
										& \textbf{}                                                           						& \textbf{available}	& \textbf{DoS}	& \textbf{Fuzzing}	& \textbf{Targeted ID}	& \textbf{}           & \textbf{}           & \textbf{}                            	\\ \hline
Ji et al. (2018) \cite{ji_comparative_2018}			& Unpublished dataset                                                   					& No				&			& Simulated         	& Simulated			& Simulated           &                     &                                      	\\ \hline
Dupont et al. (2019) \cite{dupont_evaluation_2019}		& Automotive Controller Area Network (CAN) Bus Intrusion Dataset v2 \cite{dupont_automotive_2019}	& Yes                   & Simulated		& Simulated			& Real \& simulated	& Simulated           & Simulated           & Diagnostic attack                    	\\
										& CAN Dataset for intrusion detection (OTIDS) \cite{lee_otids_2017}						& Yes                   & Real		& Real			& 				&                     &                     & Remote impersonation                 	\\
										& Car-Hacking Dataset for the intrusion detection \cite{seo_gids_2018}						& Yes                   &			&				& Real			&                     &                     &                                      	\\ \hline
Stabili et al. (2021) \cite{stabili_benchmark_2021}		& VTC2019 Dataset$^{\mathrm{a}}$ \cite{stabili_detection_2019}							& Yes				& Simulated		& Simulated			& Simulated			& Simulated           &                     & Sequence replay                      	\\ \hline
Agbaje et al. (2022) \cite{agbaje_framework_2022}		& Car-Hacking Dataset for the intrusion detection \cite{seo_gids_2018}						& Yes                   & Real		& Real			& Real			&                     &                     &                                      	\\ \hline
Blevins et al. (2021) \cite{blevins_time-based_2021}		& Real ORNL Automotive Dynamometer (ROAD) CAN Intrusion Dataset \cite{verma_addressing_2020}       	& Yes                   &			& Real			& Real			&                     &                     &                                      	\\ \hline
Stachowski et al. (2019) \cite{stachowski_assessment_2019}	& Online evaluation														& N/A				&			&				& Real			&			    & 			  & 							\\ \hline
Taylor et al. (2018) \cite{taylor_probing_2018}             & Unpublished dataset                                                   					& No                    &			& Simulated			& Simulated			&                     &                     & Sequence replay                      	\\ \hline
Berger et al. (2019) \cite{berger_comparative_2019}         & Renault Zoe datatset$^{\mathrm{b}}$ \cite{rieke_behavior_2017}   						& Yes                   &			&				& 				&                     &                     &                                      	\\ 
										& Car-Hacking Dataset for the intrusion detection \cite{seo_gids_2018}              			& Yes                   & Real		& Real			& Real			&                     &                     &                                      	\\ \hline
Moulahi et al. (2021) \cite{moulahi_comparative_2021}		& CAN Dataset for intrusion detection (OTIDS) \cite{lee_otids_2017}						& Yes                   & Real		& Real              	&				&                     &                     & Remote impersonation                 	\\ \hline
Costa Cañones (2021) \cite{costa_canones_benchmarking_2021}	& ReCAN – Dataset for reverse engineering of Controller Area Networks \cite{zago_recan_2020}		& Yes$^{\mathrm{c}}$	& 			& Simulated         	&				&                     &                     & Sequence replay; Adversarial attacks	\\ \hline
Swessi and Idoudi (2022) \cite{swessi_comparative_2022}	& Survival Analysis Dataset for automobile IDS \cite{han_anomaly_2018}						& Yes                   &			& Real              	&				&                     &                     &                                      	\\
										& Car-Hacking Dataset for the intrusion detection \cite{seo_gids_2018}						& Yes                   &     	   	& Real	           	&				&                     &                     &                                      	\\
										& Real ORNL Automotive Dynamometer (ROAD) CAN Intrusion Dataset \cite{verma_addressing_2020}		& Yes                   &          		& Real              	&				&                     &                     &                                      	\\ \hline
Anyanwu et al. (2021) \cite{anyanwu_countering_2021}		& Intrusion Detection in CAN bus Dataset \cite{sami_intrusion_2019}						& Yes                   & Simulated    	& Real \& simulated 	&				&                     &                     &                                      	\\ \hline
Okokpujie et al. (2021) \cite{okokpujie_anomaly-based_2021}	& Car Hacking Attack and Defense Challenge 2020 \cite{kang_car_2021}						& Yes                   & Real         	& Real              	& Real			&                     &                     & Replay                               	\\ \hline
\multicolumn{9}{l}{$^{\mathrm{a}}$Only the clean dataset has been used in \cite{stabili_benchmark_2021} to simulate attacks. The attack datasets available are not used in \cite{stabili_benchmark_2021}}										\\								
\multicolumn{9}{l}{$^{\mathrm{b}}$This dataset consists of only non-anomalous data and has been created for behaviour analysis. In \cite{berger_comparative_2019}, it has been used to evaluate the unsupervised learning methods only.}					\\
\multicolumn{9}{l}{$^{\mathrm{c}}$Only benign dataset available}																		
\end{tabular}%
}
\label{table-attacks}
\end{table}

\begin{table}[]
\caption{Evaluation Metrics Reported in Surveyed CAN IDS Benchmark Frameworks and Comparative Studies}
\resizebox{\linewidth}{!}{%
\begin{tabular}{@{\extracolsep{4pt}}lccccccccccp{0.12\linewidth}@{}}
\hline
\textbf{Work}								& \multicolumn{5}{c}{\textbf{Basic Security metrics}}													& \multicolumn{3}{c}{\textbf{Composite Security metrics}}							& \multicolumn{2}{c}{\textbf{Performance metrics}}	& {\textbf{Others}}                      			\\ \cline{2-6} \cline{7-9} \cline{10-11}
										& \textbf{Accuracy}	& \textbf{Precision}	& \textbf{Recall}	& \textbf{Confusion}		& \textbf{FPR$^{\mathrm{a}}$}	& \textbf{F1-score}	& \textbf{ROC$^{\mathrm{b}}$ curve}	& \textbf{P-R$^{\mathrm{c}}$ curve}	& \textbf{Training} 		& \textbf{Testing}	&                      						\\ 
										&				&				&			& \textbf{matrix}			&					&				&						&						& \textbf{time}			& \textbf{time}		&									\\ \hline
Ji et al. (2018) \cite{ji_comparative_2018}			&                   	& \checkmark		&                	& 					& \checkmark   			& \checkmark       	&\checkmark   				& \checkmark       			&                       	&				& True positive rate						\\ \hline
Dupont et al. (2019) \cite{dupont_evaluation_2019}		&                   	&                 	&			& 					& \checkmark   			&                 	&          					&                    			&                        	&				& Whether at least one alert is raised			\\ \hline
Stabili et al. (2021) \cite{stabili_benchmark_2021}		&				&                    	&			& 					&              			& \checkmark       	&             				&                    			&                        	&				& 									\\ \hline
Agbaje et al. (2022) \cite{agbaje_framework_2022}		& \checkmark        	& \checkmark         	& \checkmark	& 					& \checkmark   			& \checkmark       	&              				&                    			&                       	&                       & 									\\ \hline
Blevins et al. (2021) \cite{blevins_time-based_2021}		&                  	&                    	&			&  					&              			&                 	&              				& \checkmark         			& 	                     	&                       &									\\ \hline
Stachowski et al. (2019) \cite{stachowski_assessment_2019}	&				& \checkmark		& \checkmark	& 					& \checkmark			& \checkmark		& \checkmark				&						&					&				& False negative rate, informedness, markedness		\\ \hline
Taylor et al. (2018) \cite{taylor_probing_2018}			&                   	&                    	&                	&  					&              			&                		& \checkmark          			&                    			&                        	&                       &									\\ \hline
Berger et al. (2019) \cite{berger_comparative_2019}		& \checkmark        	&                    	&			& \checkmark$^{\mathrm{d}}$	&              			&               		& 						&                    			&                   		&                       &									\\ \hline
Moulahi et al. (2021) \cite{moulahi_comparative_2021}		& \checkmark        	& \checkmark		& \checkmark     	& \checkmark			&              			& \checkmark        	& 						&                    			&  					&                      	& 									\\ \hline
Costa Cañones (2021) \cite{costa_canones_benchmarking_2021}	& \checkmark        	& \checkmark		& \checkmark	&					&              			& \checkmark        	& 						&                    			& 					&                      	& False negative rate (only for adversarial attacks)	\\ \hline
Swessi and Idoudi (2022) \cite{swessi_comparative_2022}	& \checkmark        	&                    	&			& 					& \checkmark   			& \checkmark        	& \checkmark          			& \checkmark         			& \checkmark			& \checkmark           	& Balanced accuracy						\\ \hline
Anyanwu et al. (2021) \cite{anyanwu_countering_2021}		& \checkmark        	&                    	& 			& \checkmark$^{\mathrm{d}}$	&              			&                 	& \checkmark$^{\mathrm{d}}$		&                    			& \checkmark$^{\mathrm{d}}$	&                       & Minimum classification error $^{\mathrm{d}}$		\\ \hline
Okokpujie et al. (2021) \cite{okokpujie_anomaly-based_2021}	& \checkmark        	& \checkmark		& \checkmark	& \checkmark			&              			& \checkmark        	&              				&                    			&                        	&                       &                                     			\\ \hline
\multicolumn{12}{l}{$^{\mathrm{a}}$False positive rate}					\\
\multicolumn{12}{l}{$^{\mathrm{b}}$Receiver Operator Characteristic}			\\  
\multicolumn{12}{l}{$^{\mathrm{c}}$Precision-Recall}						\\
\multicolumn{12}{l}{$^{\mathrm{d}}$Only for selected IDS under test}
\end{tabular}%
}
\label{table-metrics}
\end{table}

\end{landscape}
} 

\subsection{Statistical intrusion detection methods}

\textbf{Ji et al.} \cite{ji_comparative_2018} is one of the first works to present a comparative study of light-weight, statistical CAN intrusion detection methods from the literature. Four statistical-based intrusion detection methods are evaluated, which are based on analysing information entropy, clock skew, ID sequences, and CAN bus throughput, respectively. These methods use only AID and arrival timestamps as features. Four simulated attack datasets have been used for evaluation, which, like many other attack datasets, are highly imbalanced. The clock skew approach was found to be the best overall across the tested attack scenarios, with almost perfect true positive rates (TPR) and false positive rates (FPR). Among the other methods, the throughput approach was best at detecting flooding, while the ID sequences method was good at detecting the injection of forged messages. However, the entropy approach did not yield good results for replay attacks, where the entropy of the CAN bus stream is not significantly changed.

\textbf{Dupont et al.} \cite{dupont_evaluation_2019} present a unified framework for CAN IDS evaluation as well as a publicly available dataset created for the same. This includes benign data collected from two live vehicles and a CAN bus prototype. Various fabrication attacks as well as a suspension attack are simulated on the benign dataset to create the attack dataset (except for one real targeted ID attack on the prototype). Two other publicly available datasets consisting of real attacks have also been used. It was found that most of the methods are only able to detect attacks that cause drastic changes, as in the case with flooding. The methods are described as relying on narrow indicators of compromise and producing too many false positives. The authors suggest that content-aware methods that take into account not just the bit representation but also the semantics of CAN messages would yield better results. 

The benchmark framework by \textbf{Stabili et al.}  \cite{stabili_benchmark_2021} allows the evaluation of four IDS algorithms from the literature against a threat model consisting of three attack types: replay, fuzzing, and disruption. The attack datasets used for testing, representing seven attack scenarios, have been created by simulating attacks on a real benign dataset collected from a vehicle. The four algorithms chosen use different features of CAN bus traffic for anomaly detection: entropy, AID sequences, payloads, and timing. It was found that while the message sequence algorithm showed efficacy in all attack scenarios, the other IDS were effective only for certain attacks.

\textbf{Agbaje et al.} \cite{agbaje_framework_2022} identify inconsistencies in three aspects that hinder comparative evaluation of CAN IDS: disparate training datasets, disparate evaluation datasets, and disparate evaluation metrics. To address these inconsistencies, a flexible evaluation framework is provided to enable consistent and repeatable evaluation of CAN IDS. This work differs from previous art in that it allows the addition of new datasets and algorithms alongside the ones that have been compared in this work. A variety of intrusion detection methods have been evaluated that differ in the features and techniques used. It is observed that different methods have their own strengths and weaknesses and are not equally capable of detecting all attacks. The authors conclude that more generalised methods that take into account the interrelationships between messages, such as graph- and ML-based methods, could be better suited to detect a large variety of attacks. 

The study by \textbf{Blevins et al.} \cite{blevins_time-based_2021} uniquely focuses on benchmarking only timing-based intrusion detection methods against the ROAD dataset \cite{verma_addressing_2020}, which consists of real attacks with verified physical effects. Alongside fuzzing attacks, the dataset also includes targeted ID attacks using flam delivery, which makes them stealthy attacks with minimum numbers of injected messages. Four statistical methods utilizing timing of CAN messages have been evaluated against fuzzy attack and targeted ID attack logs. The binning and mean inter-message time methods performed well, both with and without outliers, as opposed to the methods relying on fitting a distribution curve to the data. Binning was the best detector in terms of both the area under curve (AUC) of the PR-curve and the F1-score, and was implemented in an OBD-II plug-in prototype. While the experiments reported here are offline experiments, a detection latency analysis is provided for the binning algorithm.

\textbf{Stachowski et al.} \cite{stachowski_assessment_2019} provide an assessment methodology for evaluating and comparing intrusion detection products designed for the automotive CAN bus. This evaluation methodology differs from the rest of the surveyed works in that it involves online evaluation of CAN IDS---all the CAN IDS being evaluated are integrated into test vehicles for real-time intrusion detection. Three anomaly-based CAN IDS products have been evaluated to demonstrate this methodology, but further details on the vendors or intrusion detection methods have not been provided. While the methodology includes qualitative and quantitative metrics, the assessment carried out used only quantitative metrics related to attack detection accuracy. A large number of targeted ID attacks were carried out on the CAN bus of test vehicles while the vehicles were stationary and in motion. The IDS were also tested in driving scenarios with no attacks being carried out. Furthermore, evaluations were carried out in two phases, with IDS vendors given the opportunity to fine-tune their IDS for the second phase. As a result, the evaluated CAN IDS were generally found to perform better in the second phase, with higher true positive rates (TPR) and lower false positive rates (FPR). However, none of the IDS evaluated were found to be effective in detecting all attacks.

\subsection{ML-based intrusion detection methods}

\textbf{Taylor et al.} \cite{taylor_probing_2018} develop IDS based on two types of recurrent neural networks (RNN)---long short-term memory (LSTM) and gated recurrent units (GRU)---and compare them against Markov models. To facilitate IDS evaluation, a comprehensive attack framework is proposed with parametrised attack descriptions that can be used to generate realistic, representative attack simulations. This attack framework has been used to create attack datasets for 537 test cases with different types of fuzzing and targeted ID attacks. It is found that the RNN models generally perform very well with high AUC measures, with the large LSTM model deemed the best. On the other hand, Markov models were not much better than chance at anomaly detection. The effect of changing attack parameters has also been reported, whereby the performance of the LSTM detection model is found to vary depending on the variability of the CAN signals, AIDs, and attack types.

\textbf{Berger et al.} \cite{berger_comparative_2019} examine the performance of four supervised and unsupervised learning methods against the same datasets containing DoS, fuzzing, and targeted ID attacks. Experiments were performed by varying the number of training samples (for OCSVM and SVM) and the number of neurons (for neural networks and LSTM). OCSVM demonstrated a high bias towards predicting the normal class. On the other hand, SVM and neural network provided high accuracies, but require attack samples for training,  which is disadvantageous. Although LSTM demonstrated diminished performance and is the most computationally intensive method tested, the authors conclude that it can be improved and is a viable method of intrusion detection.

The comparative study of \textbf{Moulahi et al.} \cite{moulahi_comparative_2021} differs from the works discussed thus far in that the IDS models implemented are multi-class classification models with three attack classes. Four traditional classification algorithms have been evaluated with a feature set that not only includes the AID and payload data of each CAN message but also the AIDs of the previous three messages. While the remote impersonation attack is detected with near-perfect metrics, the performances of the evaluated models suffer when it comes to the fuzzing and DoS attacks, due to the lower number of examples of the latter attack types in the dataset. The findings of this work have also been put into context by comparing the results with previous work using the same dataset. Overall, the random forest (RF) model in this work is found to be the best in terms of both detection accuracy and training and testing times. 

The benchmark framework provided by \textbf{Costa Cañones} \cite{costa_canones_benchmarking_2021} evaluates Isolation Forest, OCSVM, and autoencoder intrusion detection models against various simulated attack datasets. Apart from three basic attack scenarios, this work uniquely presents two sophisticated adversarial attack scenarios to examine IDS evasion. The IF and OCVSVM were shown to have inferior performance compared to the autoencoder detectors, particularly in terms of accuracy and recall. Furthermore, it is shown that the autoencoder detectors that also took into consideration time sequences were effective for detecting attacks with valid payloads. They were also more resilient against adversarial attacks designed to evade them. 

Like in \cite{blevins_time-based_2021}, \textbf{Swessi \& Idoudi} \cite{swessi_comparative_2022} study a particular type of intrusion detection models---ensemble learning---for the detection of fuzzing attacks. Fuzzing attacks can be difficult to detect, especially when the injected messages use legitimate AIDs. This extensive study compares 11 different ensemble algorithms against three real fuzzing attack datasets. Ensemble learning methods were generally found to be very effective with very high accuracies, but bagging, extreme Gradient Boosting (XGB), light gradient boosting (LGB), and category gradient boosting (CGB) were the best in terms of both detection rates as well as training and testing times. 

\textbf{Anyanwu et al.} \cite{anyanwu_countering_2021} conducted a comparative study of a total of 22 different ML models across six types, to determine the best algorithm for CAN intrusion detection. Although the attack model or attacks tested are not described in this work, the documentation of the datasets used for evaluation describes them as containing fuzzing and DoS attacks. Decision trees, KNN, SVM, and ensemble models gave accuracies of 100\%. Training time and the number of misclassifications were used to differentiate among these models with perfect accuracies; the decision tree classifier of the type Fine was deemed the best considering all these factors. 

\textbf{Okokpujie et al.} \cite{okokpujie_anomaly-based_2021} also conducted a comparative study between Feedforward Neural Network (FNN) and SVM models against four different real attack datasets. The SVM models evaluated include those with linear, polynomial, radial basis, and signmoid kernels. As in \cite{moulahi_comparative_2021}, intrusion detection has been modeled as a multi-class classification problem that includes attack type classification. The authors note that accuracy in itself is not a sufficient measure of detection performance, particularly with unbalanced attack datasets, and therefore also report precision, recall, and F1-score. Considering all reported metrics, SVM with the radial basis kernel was found to be the best, while the FNN model was not able to detect some of the attack types at all.

\section{Discussion}\label{s6_discussion}

In this section, the reviewed benchmarking framework and comparative studies are further categorised and discussed in terms of the proposed design space to understand current efforts as well as opportunities for future work. 

\subsection{IDS type}

Regarding the type of CAN IDS, anomaly-based intrusion detection methods are clearly seen as the way forward for CAN intrusion detection---anomaly-based methods are the only type of IDS that were found to have been benchmarked and compared in the surveyed works. In the wider CAN IDS literature, anomaly-based IDS are indeed the most common type of IDS proposed \cite{aliwa_cyberattacks_2021, karopoulos_demystifying_2022}. Signature-based methods are not very common since complete CAN attack signatures do not exist, and such databases are difficult to create as the implementation of CAN messaging differs among vehicles of different models and makes \cite{aliwa_cyberattacks_2021}.

Among the anomaly-based methods, each of the surveyed studies restricted themselves to either one of two types of IDS: statistical methods or ML-based methods. The exception to this is the evaluation framework in \cite{agbaje_framework_2022}, where a neural network IDS is included along with the other statistical methods. Papers benchmarking statistical methods \cite{agbaje_framework_2022, stabili_benchmark_2021, dupont_evaluation_2019, ji_comparative_2018} include a variety of IDS that use different types of features and techniques for attack detection, including timing and frequencies of AIDs, AID sequences, Hamming distances in payload, and entropy of CAN messages. Meanwhile, studies of ML-based methods have applied algorithms like SVM, decision trees, Isolation Forest, ensemble algorithms, as well as various types of neural networks. 

Since intrusion detection methods are not equally effective at detecting all attack types, some studies focus on particular intrusion detection techniques or attack types. Blevins et al. \cite{blevins_time-based_2021} benchmark only timing-based intrusion detection methods, which are computationally inexpensive but are effective only against attacks that alter the timing of CAN messages on the CAN bus. This is why only fuzzing and targeted ID attacks are used for evaluation and not masquerade attacks, which do not impact message timing. Another work \cite{swessi_comparative_2022} examines the detection of fuzzing attacks in particular, which can be difficult to detect, using ensemble learning algorithms that use timestamps, AIDs, and payload data as features.


It is also observed that while most of the surveyed works treat anomaly detection as a binary classification problem, two of these---\cite{moulahi_comparative_2021} and \cite{okokpujie_anomaly-based_2021}---model intrusion detection as a multi-class classification problem aiming to classify attack data into attack types. Attack classification can indeed be a useful addition to an IDS since the type of attack can inform attack mitigation responses. 

Finally, we find no benchmarking or comparative evaluation study that includes physical characteristics-based CAN IDS. Review of physical layer CAN IDS reveals that they are commonly evaluated in online tests with testbeds and real vehicles. These evaluation methods are not common in the surveyed works, which take advantage of publicly available datasets. This indicates a need for comparative evaluation frameworks for online testing as well as datasets (such as \cite{foruhandeh_simple_2019, popa_ecuprintphysical_2022}), which would encourage similar benchmarking studies of physical layer CAN IDS.

\subsection{Evaluation type}\label{designspace_evaluationtype}

If we distinguish between offline and online evaluations, almost all of the works surveyed have performed benchmarking with offline evaluations, whereby the assessed CAN IDS analyse collected CAN bus datasets. This can be explained by the proliferation of publicly available datasets covering common attack types in the literature, such as fabrication attacks. With comprehensive documentation, offline experiments using datasets can be replicated easily. Offline assessments can also be conveniently used to evaluate multiple IDS on an equal footing against the same dataset under equivalent test conditions. 

On the other hand, the assessment methodology in \cite{stachowski_assessment_2019} is the only one among the surveyed works to include online evaluations of CAN IDS, whereby IDS products have been integrated into test vehicles for evaluation in different scenarios. As noted in Section \ref{designspace_evaluationtype}, online tests are relatively difficult to perform with a real vehicle due to costs and safety risks. Notably, we did not find any studies that utilize testbeds (such as that proposed in \cite{jadidbonab_real-time_2021}) or simulations (such as Vector CANoe, as used in \cite{ujiie_method_2016}) for the purpose of benchmarking, both of which are options that provide a more realistic environment for assessing detection capability and performance without the disadvantages of using real vehicles. This indicates a need for mature frameworks for real-time assessment and benchmarking of CAN IDS that can not only be implemented without the costs and risks associated with test vehicles but also allow repeatable assessments of CAN IDS under identical experimental conditions. 

\subsection{Workload}

As mentioned in the previous subsection, the most common method used for comparative evaluation of CAN IDS is using datasets for offline experiments. The popularity of real CAN bus datasets in the broader CAN literature is reflected among the surveyed benchmarking frameworks and comparative studies, wherein all the datasets used are derived from the CAN bus of real vehicles. However, they differ in whether the attacks themselves are real or simulated. Half of the works surveyed use publicly available datasets from HCRL \cite{lee_otids_2017, seo_gids_2018, han_anomaly_2018, kang_car_2021}, which consist of real attack traces. Another dataset including real attacks with physically verified effects is the ROAD dataset \cite{verma_addressing_2020}, which was more recently published and has been used in only two of the surveyed works \cite{blevins_time-based_2021, swessi_comparative_2022}. Other than these, the attack datasets that are used have been created by modifying logs of real CAN bus traffic to simulate attacks on the CAN bus. One of the earliest works \cite{ji_comparative_2018} performed their comparative evaluation with an unpublished dataset whereby benign data collected from a real car is replayed in a CAN bus simulation software, where attack scenarios are conducted to generate attack samples. The study by Taylor et al. \cite{taylor_probing_2018} is unique in that they propose a method for generating attack samples using a parametrized attack framework. While it may be argued that approaches such as these \cite{ji_comparative_2018, taylor_probing_2018} may not produce realistic attack samples, these methods provide a customisable way of generating attack samples, enabling assessment against a variety of attack scenarios---from high frequency message injection attacks to low rate injection attacks and masquerade attacks. 
 
Unlike the remaining studies, \cite{stachowski_assessment_2019} perform online evaluations on real test vehicles, the CAN buses of which are made to undergo different types of targeted ID attacks generated using attack scripts.

\subsection{Attack model}\label{subsec-attacks-tested}

Among all the works surveyed, fabrication attacks are the most common attack type tested, which is due to the fact that fabrication attacks comprise the majority of attacks found in CAN IDS literature, and there are several CAN datasets that provide real CAN bus logs with real attack samples \cite{verma_addressing_2020}. On the other hand, few works evaluate intrusion detection methods against suspension and masquerade attacks. Only one study \cite{dupont_evaluation_2019} performed IDS evaluations against all attack types in our attack model, but with simulated attack datasets. This highlights the need for realistic samples of these attack types.

The benefit of using real attacks over simulated attacks is realised in the online assessments in \cite{stachowski_assessment_2019}, where the physical effects of attacks like doors locking could be observed and verified and the evaluated CAN IDS could be confirmed to be effective in realistic attack scenarios. This is also the case for studies that use datasets like \cite{verma_addressing_2020, seo_gids_2018} that were collected from a real vehicle undergoing cyberattacks. However, real attack datasets that have been used in the surveyed works have their limitations. For instance, the dataset \cite{seo_gids_2018} used in \cite{agbaje_framework_2022, dupont_evaluation_2019} does not provide a difficult detection challenge since it consists of fabrication attacks that can be easily detected by trivial IDS methods and may not sufficiently test IDS methods that are capable of detecting more subtle attacks \cite{verma_addressing_2020}. 

Apart from the attack classes identified in Section \ref{can_attack_model}, ML-based CAN IDS have also been demonstrated to be vulnerable against adversarial attacks \cite{choi_robustness_2021} that are designed to evade detection. The benchmark framework in \cite{costa_canones_benchmarking_2021} also evaluates the IDS under test against adversarial attack samples generated using Generative Adversarial Networks (GAN) and heuristics.

\subsection{Evaluation metrics}

From the evaluation metrics described in Section \ref{designspace_evaluationmetrics}, we find that the surveyed studies all focus on examining attack detection accuracy by presenting security-related metrics almost exclusively. These metrics include classification accuracy, precision, recall, false positive rate (FPR), and receiver operator characteristic (ROC) curve. In order to avoid the pitfalls associated with relying on only a single metric like classification accuracy \cite{stakhanova_analysis_2017}, the majority of the works report multiple metrics to provide a complete picture of attack detection capability. The study in \cite{swessi_comparative_2022} also reports balanced accuracy in addition to other metrics to account for class imbalance in attack datasets. Furthermore, as noted in subsection \ref{subsec-attacks-tested}, resistance to evasive attacks is evaluated in the benchmark framework provided in \cite{costa_canones_benchmarking_2021}.

On the contrary, performance metrics are not widely reported in the surveyed studies. None of the surveyed works report detection latency, with only one work \cite{blevins_time-based_2021} providing an analysis of detection latency in terms of the computational time of the detection algorithms. Detection speed and latency to filter benign messages are included in the suite of metrics in the online assessment by Stachowski et al. \cite{stachowski_assessment_2019}, but they are ultimately not recorded and reported. Another study \cite{swessi_comparative_2022} reports testing execution time in addition to training time as a means of indicating performance. This trend can be explained by the fact that, unlike the security-related metrics that can be obtained in offline experiments with datasets, obtaining precise measurements of detection latency requires some form of online testing, such as simulations or testbeds. This further highlights the necessity of online testing for benchmarking CAN IDS. We also find a limited examination of non-functional properties such as resource consumption, performance overhead, and workload processing capacity. 

\section{Recommendations for future work}\label{s7_futurework}

In compiling our evaluation design space and conducting the survey of benchmarking and comparative evaluation studies, we found several concerns that lead to opportunities for future work. First, we observed a lack of studies that incorporate CAN IDS from different categories. Only \cite{agbaje_framework_2022} and \cite{dupont_automotive_2019} include an ML-based and signature-based CAN IDS, respectively, in their studies of otherwise statistical CAN IDS. We were also not able to find studies that benchmark physical layer CAN IDS along with data link layer CAN IDS, or indeed, a study with only physical layer CAN IDS. Secondly, offline evaluations with CAN datasets were found to be the most prevalent methodology employed for benchmarking, with only one study \cite{stachowski_assessment_2019} performing CAN IDS evaluations on real vehicular networks. However, this study focuses on presenting an assessment framework and does not disclose the mechanics of the CAN IDS being evaluated, thus providing no insight into detection techniques and their effectiveness against the selected attack types. We also observe that more complex attack types like suspension and masquerade are not considered in the surveyed works, while in the broader CAN IDS literature, there are numerous CAN IDS that consider the detection of masquerade attacks \cite{moriano_detecting_2022, nichelini_canova_2023, zhang_hybrid_2022}. Finally, we find that all the reviewed studies are restricted to the assessment of security-related properties, and most do not provide assessment of performance factors such as detection latency or resource overhead. In light of these issues, we \rev{provide} the following recommendations for future work:

\paragraph{Comprehensive benchmarking datasets} Benchmarking datasets collected from real vehicles and consisting of real attack traces are crucial for the development and evaluation of CAN IDS. As opposed to synthetic datasets and simulated attacks, real traces with real attacks capture the dynamics of CAN bus traffic and are best suited for evaluating CAN IDS. While a number of CAN datasets have been published for IDS evaluation \cite{karopoulos_demystifying_2022, swessi_comparative_review_datasets_2021, verma_addressing_2020}, none has emerged as a benchmark dataset in the way the KDD Cup and DARPA datasets are used for computer network IDS evaluation \cite{aliwa_cyberattacks_2021}. There is also a need for comprehensive datasets of real attack traces of all known types. While several traces containing fabrication attacks are available, currently available datasets do not contain real traces of suspension or masquerading attacks, and more attacks are being discovered. The creation of such a dataset would allow robust benchmarking and evaluation of CAN IDS, ensuring that CAN IDS being developed can effectively detect at least all known attacks. 

Methods of synthesizing datasets and simulating attacks are also not without merit: creating real attack datasets requires considerable resources and skill, and simulation tools can allow the creation of customisable attacks, as the parametrised attack framework in \cite{taylor_probing_2018} illustrates. Such datasets can therefore fill in gaps where real datasets are not available. 

\paragraph{Online evaluation methods} Current benchmarking and evaluation studies mostly use offline methods with CAN bus datasets. This restricts the kinds of CAN IDS properties that can be assessed -- the surveyed works almost exclusively focus on evaluating attack detection accuracy. Apart from attack detection accuracy, attack detection latency, resource consumption, and workload processing capability are important considerations for an in-vehicle IDS but have not been examined sufficiently in current CAN IDS benchmarking studies. Therefore, there is a need for online evaluation methods for CAN IDS, such as in \cite{stachowski_assessment_2019}, that can allow repeatable comparative evaluation studies. Benchmarking with online testing methods, such as using trace replay tools, simulations, and testbeds, can facilitate not just the assessment of non-functional properties but also allow a more accurate evaluation of attack detection capability in a manner close to real operating environments. Documentation becomes an important aspect of reporting online evaluation conducted using real vehicles, testbeds, and simulations; documenting and reporting hardware devices and components, network topologies, and source code used would further enhance the reproducibility of results.

\paragraph{Benchmarking Layer 1 CAN IDS} While most CAN intrusion detection methods in the literature use features derived from OSI Layer 2 data, i.e., CAN frames, the physical characteristics-based methods use Layer 1 information for attack detection. So current CAN attack datasets cannot be used to evaluate these physical IDS and, as a result, these IDS have not been benchmarked alongside the CAN IDS using CAN messages. Hence, inclusive benchmarking methods are needed where both Layer 1, Layer 2, and even hybrid CAN IDS can be evaluated and compared with each other under similar test conditions. Apart from using real vehicles and testbeds, this can be realised by creating datasets of physical measurements (such as \cite{foruhandeh_simple_2019, popa_ecuprintphysical_2022}) and datasets with both Layer 1 and Layer 2 data for offline testing, as well as simulations of physical characteristics for online testing. Inclusive benchmarking of Layer 1 and Layer 2 CAN IDS will allow direct comparison among different types of CAN IDS and also facilitate the development of comprehensive intrusion detection methods that incorporate more diverse features and detect a wider range of attacks.

\paragraph{Comprehensive evaluation metrics} Among the surveyed studies, it is observed that the set of metrics reported differs, which hinders direct comparison across studies. Furthermore, the focus is only on security-related metrics, with performance-related metrics largely not being measured or reported. But in order to select an IDS for implementation in automotive networks, it is necessary to assess not just attack detection capability but also detection latency and other non-functional properties. This means that a comprehensive suite of evaluation metrics needs to be developed that includes both security- and performance-related metrics, covering the assessment of CAN IDS in all practical aspects. The selection of security-related metrics should consider factors such as class imbalance in attack datasets and prior likelihood of attack, which make using only classification accuracy or ROC insufficient \cite{stakhanova_analysis_2017} and necessitate metrics like balanced accuracy and MCC, which give equal importance to both normal and attack classes. 

\section{Conclusion}\label{s8_conclusion}

Many intrusion detection methods are being developed for the CAN bus in an endeavour to secure it against various types of cyberattacks that have the potential to cause vehicles to malfunction and result in dangerous accidents. The evaluation of these CAN IDS varies in terms of the CAN IDS type assessed, the attack types considered, the evaluation type, the workload used for evaluation, and the evaluation metrics reported. We thus propose a CAN IDS evaluation design space in the manner of \cite{milenkoski_evaluating_2015} encapsulating these five aspects of CAN IDS assessment, with the aim of categorizing current CAN IDS works and serving as a guide for planning evaluation studies by enumerating existing approaches to CAN IDS evaluation. 

CAN IDS are usually evaluated under disparate experimental conditions, which hinders direct comparison. Therefore, there have been a number of benchmark frameworks proposed and comparative studies conducted that evaluate CAN IDS in similar experimental conditions to reveal how they perform in relation to each other. Such benchmarking efforts ultimately facilitate the selection of the most appropriate CAN intrusion detection methods for implementation in in-vehicle networks. This work surveys current efforts at benchmarking and comparing CAN IDS and discusses them in terms of the proposed CAN IDS evaluation design space in order to understand current trends as well as directions for future work. 

From the surveyed works, it is apparent that anomaly-based CAN IDS are the most popular type of CAN IDS selected for benchmarking since they have the capability to detect novel, unknown attacks and do not require attack signatures. Among anomaly-based CAN IDS, it is observed that only statistical- and ML-based methods are the ones that are typically included in benchmarking studies. Because of the difficulties associated with conducting online tests, all comparative evaluations are offline evaluations using CAN bus datasets. There are a number of publicly available traces of CAN bus traffic collected from real vehicles, both under normal operation and under attack, which are commonly used for offline evaluations. However, such datasets are limited in the types of attacks they contain; while there are several datasets available with common fabrication attacks, there is a lack of datasets containing other classes of attacks like suspension and masquerade attacks. Offline experiments also allow measurement of only security-related metrics related to attack detection accuracy. As such, attack detection latency and other non-functional properties are understudied in current benchmarking and comparative studies. 

Examining surveyed works in terms of this design space reveals avenues for future work: benchmarking datasets, repeatable online evaluations, methods for comparing  Layer 1 CAN IDS with Layer 2 CAN IDS, and comprehensive evaluation metrics.  





\bibliographystyle{ios1}           
\bibliography{bibliography}        

\end{document}